\documentclass[aps,pra,twocolumn, showkeys,superscriptaddress]{revtex4}
\usepackage{amsmath,paralist,amsthm,comment,amssymb}

\DeclareMathOperator{\wt}{wt}

\newcommand{\mc}[1]{\mathcal{#1}}
\newcommand{\C}{\mathbb{C}}
\newcommand{\F}{\mathbb{F}}
\newcommand{\nix}[1]{}
\newcommand{\floor}[1]{{\left\lfloor #1\right\rfloor}}
\newcommand{\ceil}[1]{{\left\lceil #1\right\rceil}}

\newtheorem{theorem}{Theorem}
\newtheorem{corollary}[theorem]{Corollary}
\newtheorem{lemma}[theorem]{Lemma}
\newtheorem{proposition}[theorem]{Proposition}

\begin{document}

\title{Degenerate Quantum Codes  and the Quantum Hamming Bound}

\author{Pradeep Sarvepalli}
\email[]{pradeep@phas.ubc.ca}
\affiliation{Department of Physics and Astronomy, University of British Columbia, Vancouver V6T 1Z1, Canada }

\author{Andreas Klappenecker}
\email[]{klappi@cse.tamu.edu}
\affiliation {Department of Computer Science, Texas A\&M University, College  Station, TX 77843}


\date{November 21, 2009}

\begin{abstract}
The parameters of a nondegenerate quantum code must obey the Hamming
bound. An important open problem in quantum coding theory is whether
or not the parameters of a degenerate quantum code can violate this
bound for nondegenerate quantum codes.  In this paper we show that  
 Calderbank-Shor-Steane (CSS) codes with alphabet $q\geq 5$ cannot beat
the quantum Hamming bound. We prove a quantum version of the Griesmer
bound for the CSS codes which allows us to strengthen the Rains' bound that
an $[[n,k,d]]_2$ code cannot correct more than $\floor{(n+1)/6}$ errors
to $\floor{(n-k+1)/6}$.  Additionally, we also show that the general 
quantum codes $[[n,k,d]]_q$ with $k+d\leq {(1-2eq^{-2})n}$ cannot beat the 
quantum Hamming bound. 

\end{abstract}


\maketitle
\section{Introduction}
Quantum information can be protected by encoding it into a quantum
error-correcting code.  An $((n,K,d))_q$ quantum code is a
$K$-dimensional subspace of the state space $\mc{H}=(\C^{q})^{\otimes
n}$ of $n$ quantum systems with $q$ levels that can detect all errors
affecting less than $d$ quantum systems, but cannot detect some errors
affecting $d$ quantum systems. An $((n,K,d))_q$ quantum code with
$k=\log_q K$ is also said to be an $[[n,k,d]]_q$ quantum code. The 
parameter $k$ is not necessarily integral.

A measure of the performance of the quantum code is its ability to
correct errors on the encoded information.  Let $\mc{B}(\mc{H})$
denote the algebra of bounded operators on~$\mc{H}$. We denote by
$P_{\mc{Q}}$  the orthogonal projector in $\mc{B}(\mc{H})$ that projects 
onto the
quantum code $\mc{Q}$.  Let $\mc{E}$ denote a subspace of
$\mc{B}(\mc{H})$ with basis $B$.  The quantum code $\mc{Q}$ is able to
correct all errors in $\mc{E}$ if and only if there exists a hermitian
matrix~$C$ such that 
\begin{equation}\label{eq:correctable} 
(P_{\mc{Q}} E^\dagger F P_{\mc{Q}})_{E,F\in
B}=C\otimes P_{\mc{Q}}.
\end{equation}
In other words, $\mc{Q}$ can correct all errors in
$\mc{E}$ if and only if it can detect all errors in the set
$\{E^\dagger F\,|\, E,F\in B\}$.

Of particular interest are localized errors that affect few quantum
systems.  Let $\mc{E}_t$ denote the vector space spanned by all
elements in $\mc{B}(\mc{H})$ affecting at most $t$ quantum systems. A
quantum code $\mc{Q}$ is called $t$-error correcting if and only if it
can correct all errors in $\mc{E}_t$. An $((n,K,d))_q$ quantum code is
$t$-error correcting for $t=\floor{
(d-1)/2}$. 

The pair $(\mc{Q},\mc{E})$ consisting of a quantum code $\mc{Q}$ and a
vector space of errors $\mc{E}$ is called degenerate if and only if
the hermitian matrix $C$ in equation~(\ref{eq:correctable}) is
singular; otherwise, $(\mc{Q},\mc{E})$ is called nondegenerate. An
$((n,K,d))_q$ quantum code $\mc{Q}$ is said to be nondegenerate if and
only if $(\mc{Q},\mc{E}_t)$ is nondegenerate for $t=\floor{ (d-1)/2}$.

In the construction of quantum codes, one would like to have both
large dimension $K$ and large minimum distance~$d$, but these are two
conflicting requirements on the quantum code. The trade off between
the number of correctable errors and the size of the quantum code is
usually quantified by various bounds. For example, a nondegenerate
$((n,K,d))_q$ quantum code satisfies the Hamming bound
\begin{eqnarray}\label{eq:qhb}
K \leq
\frac{q^n}{\sum_{j=0}^{\floor{(d-1)/2}}\binom{n}{j}(q^2-1)^j}.
\end{eqnarray}
The term `degenerate quantum code' was introduced a decade
ago. Since the term was coined, researchers raised the question
whether a degenerate $[[n,k,d]]_q$ quantum code violating the Hamming
bound (\ref{eq:qhb}) might exist, \cite{gottesman96}. The standard proof of (\ref{eq:qhb})
by a simple counting argument can fail for degenerate quantum codes in
a spectacular fashion, fueling the interest in this problem.
To  date this problem remains to be fully settled.

We review briefly some previous work to put our result in context. 
Gottesman reported the first analytical result as to the generality of
the quantum Hamming bound in \cite{gottesman97} by proving that single and double
error-correcting binary stabilizer codes cannot beat the quantum
Hamming bound. Subsequently, Ashikhmin and Litsyn \cite{ashikhmin99} showed 
a stronger result that asymptotically binary quantum codes obey the quantum Hamming
bound; their result is applicable to general codes not just binary stabilizer codes. 
In \cite{ketkar06} Gottesman's result was generalized for nonbinary 
codes with distance three \cite{ketkar06}, suggesting that even with the freedom of
increased alphabet it may not be possible to beat the quantum Hamming bound.   

In this paper we prove some new results on the applicability of
quantum Hamming bound to quantum codes.  We show that all CSS codes
with alphabet size $q\ge 5$ must obey the Hamming bound.  In the
process, we also show a weaker result that holds for general quantum
codes, namely we prove that if one bounds $k+d$ by a fraction of the
length $n$, then an arbitrary $[[n,k,d]]_q$ quantum code must also
obey the quantum Hamming bound. Furthermore, we prove a quantum
version of the Griesmer bound for the CSS codes. As a consequence of
this bound we can tighten Rains' bound when applied to CSS codes.

Since one-dimensional quantum codes are by definition nondegenerate,
hence obey the Hamming bound, we may assume throughout that the
quantum code is of dimension $K>1$.

\section{Quantum Hamming Bound and Arbitrary Quantum Codes}
One of the long standing open questions in quantum coding theory is
whether the Hamming bound (\ref{eq:qhb}) holds for degenerate quantum
codes.  In this section, we show that this question has an affirmative
answer for a large class of general quantum codes.

We denote by $h(x)=-x\log_2x-(1-x)\log_2(1-x)$ the binary
entropy function.

\begin{theorem}\label{th:qhbCondTight}
If $ 2eq^{-2} \leq \delta \leq 1$ and $q\geq 3$, then an $((n,K,d))_q$
code with $\log_q K+d \leq (1-\delta)n$ satisfies the quantum Hamming
bound~(\ref{eq:qhb}).
\end{theorem}
\begin{proof}
We have $K \leq q^{{(1-\delta)n}-d}= q^n/q^{\delta n+d}$.
Let \begin{eqnarray} T= q^{\delta n+d}\!\Big/\,
{\sum_{j=0}^t\binom{n}{j}(q^2-1)^j}. \label{eq:T0}
\end{eqnarray}
It suffices to 
show that $T\geq 1$, since this implies that 
$$ K \leq \frac{q^n}{q^{\delta n+d}} \leq \frac{q^n}
{\sum_{j=0}^t\binom{n}{j}(q^2-1)^j}.$$ As $2t+1\leq d\leq 2t+2$ we can bound $T$ from below by
\begin{eqnarray*}
T &\geq &  \frac{q^{\delta n+2t+1}}{(q^2-1)^t\sum_{j=0}^t\binom{n}{j}} = 
\frac{q^{\delta n+1}}{(1-q^{-2})^t\sum_{j=0}^t\binom{n}{j}}.
\end{eqnarray*}
By \cite[Corollary 23.6]{jukna01} 
we have $\sum_{j=0}^t\binom{n}{j} \leq 2^{nh(t/n)}$. Hence, we obtain 
\begin{eqnarray*}
T\geq \frac{q^{\delta n+1}2^{-nh(t/n)}} {(1-q^{-2})^t } = \frac{q^{\delta n + 1 -
nh(t/n)\log_q2}}{q^{t\log_q(1-q^{-2})}} \geq1. 
\end{eqnarray*}
In other words, we need to show that 
\begin{eqnarray*}
\delta n + 1 - nh(t/n) \log_q 2 -t\log _q (1-q^{-2}) \geq 0,
\end{eqnarray*}
that is 
\begin{eqnarray}
h(t/n)\log_q 2 + (t/n)\log_q(1-q^{-2}) - 1/n \le \delta \le 1. \label{eq:qhbCond}
\end{eqnarray}
Next, we will show  the above inequality holds for $\delta \geq 2eq^{-2}$.

Without loss of generality let us assume that 
$k+d= (1-\delta)n$ where $2eq^{-2}\leq \delta \leq 1$ and $k=\log_q K$.
By the quantum Singleton bound, $k+d\leq n-d+2$; so $d\leq \delta n+ 2$
and $t=\floor{(d-1)/2} \leq \floor{(\delta n+1)/2}$, hence, $t/n \leq \delta/2+1/2n$.

Let $f(x) = x - h(x/2)\log_q2 = x+(x/2)\log_q
(x/2) +(1-x/2)\log_q(1-x/2)$, for $x\in (0,2)$. The derivative of
$f(x)$ is given by
$$
f'(x) = 1 + \frac{1}{2}\log_q \frac{x}{2-x} =\frac{1}{2}\log_q\frac{q^2x}{2-x},
$$ which can be seen to satisfy $f'(x) > 0 $ for $x>2/(q^2+1)$.  Since
$\delta \geq 2eq^{-2}= 2e(1+q^{-2})/(q^2+1) > 2/(q^2+1)$, the function
$f(x)$ is increasing for $x\geq 2eq^{-2}$.  We claim that
$f(x)\geq 0$ for $x\geq 2eq^{-2}$ and $q\geq 3$. Indeed, we have
\begin{eqnarray*}
f(x) &=& x - h(x/2)\log_q2 \\
&=& x+(x/2)\log_q (x/2) +(1-x/2)\log_q(1-x/2)\\
&=& \log_q (q^2x/2)^{x/2}(1-x/2)^{1-x/2}
\geq f(2eq^{-2})\\
&=& \log_q e^{eq^{-2}}(1-eq^{-2})^{1-eq^{-2}}.
\end{eqnarray*}
Since $(1+z)\le e^z$ holds for all $z$, we have 
$(1-z) = 1/(1+z/(1-z)) \geq e^{-z/(1-z)}$; and as $eq^{-2}<1$ for $q\geq 3$ we obtain 
\begin{eqnarray*}
f(x) &\geq & \log_q e^{eq^{-2}} e^{-eq^{-2}} =  0, 
\end{eqnarray*}
as claimed. In particular, we have $\delta+1/n\ge h(\delta/2+1/2n)\log_q 2$
for $2eq^{-2}\leq \delta \leq 1$. 
The entropy function $h(x)$ is monotonically
increasing in $x$ for $x\in [0,1/2]$.  Since $t/n \leq \delta/2+1/2n$, 
for $2eq^{-2} \leq \delta \leq 1-1/n$, the
monotonicity of $h(x)$ implies that $h(t/n)\leq h(\delta/2+1/2n)$. 
If $1-1/n<\delta \leq 1$, then we observe that $1/2<\delta/2+1/2n \leq 3/4$, for 
$n\geq 2$. As  $h(x)=h(1-x)$, we have $h(1/4) \leq h(\delta/2+1/2n) < h(1/2)$.
But $t/n \leq 1/4$, by the Singleton bound, therefore again we have
$h(t/n) \leq h(\delta/2+1/2n)$. In either case we have 
$h(t/n)\log_q2 \leq h(\delta/2+1/2n)\log_q2 \leq \delta+1/n$.
Thus, $\delta$ satisfies the inequality~(\ref{eq:qhbCond});
note that $(t/n)\log_q(1-q^{-2})<0$.
If $n=1$, then $t=0$ and equation~(\ref{eq:qhbCond}) holds trivially for all 
$0\leq \delta\leq 1$. Hence, the quantum code obeys quantum
Hamming bound~(\ref{eq:qhb}).
\end{proof}

It follows from Theorem~\ref{th:qhbCondTight} that for any $\delta>0$,
an $[[n,k,d]]_q$ code with $k+d\leq {(1-\delta)n}$ obeys the quantum
Hamming bound for any alphabet size $q\geq \sqrt{2e/\delta}$.  This
suggests that it is less likely that one can find a degenerate quantum
code beating the quantum Hamming bound for larger alphabet
sizes. Indeed, if we choose a larger
alphabet size $q$, then we can choose a smaller parameter $\delta$, so 
the previous theorem rules out an even larger fraction of quantum
codes.

The following table list for a given alphabet size $q$ the fraction
$1-\delta$ of the length that bounds the sum of minimum distance $d$ and
dimension parameter $k$.

\begin{table}[htb]
\begin{center}
\begin{tabular}{|c|c|c|c|c|c|c|c|c|c|}
\hline
$q$& 3& 4& 5& 6&7& 8& 9& 10&11
\\\hline
$\delta$
&0.605
&0.340
&0.218
&0.152
&0.111
&0.085
&0.068
&0.055
&0.045
\\\hline
$1-\delta$
&0.395
&0.660
&0.782
&0.848
&0.889
&0.915
&0.932
&0.945
&0.955
\\\hline
\end{tabular}
\end{center}
\caption{Threshold values of $\delta$ for $[[n,k,d]]_q$ codes as
computed by
Theorem~\ref{th:qhbCondTight}}\label{tab:deltaRangesTight}
\end{table}

The thresholds on $\delta$ given in Theorem~\ref{th:qhbCondTight}
 are monotonically decreasing
in~$q$. Therefore, if we conclude from Theorem~\ref{th:qhbCondTight}
that all $[[n,k,d]]_\alpha$ codes
with $k+d\leq {(1-\delta)n}$ obey the Hamming bound, then this implies that the same claim holds for all alphabet sizes $q\geq \alpha$.  In particular, we can
conclude from
Table~\ref{tab:deltaRangesTight} that if $q\geq 4$ and $k+d\leq {n/2}$, 
then an $[[n,k,d]]_q$
quantum code cannot beat the quantum Hamming bound.
Similarly, we can conclude from Table~\ref{tab:deltaRangesTight}
that if $q\geq 5$ and $k+d\leq {3n/4}$, then an $[[n,k,d]]_q$  
cannot beat the quantum Hamming bound.

Notice that these results are not a restatement of the asymptotic
versions of the quantum Hamming bound. The asymptotic forms usually
claim that for large $n$, the quantum Hamming bound holds. In
contrast, the present result specifies the restriction of $K$ and $d$
when the quantum Hamming bound holds exactly, irrespective of the size
of $n$.

\section{Quantum Hamming Bound and CSS Codes}
In this section, we focus on a subset of the stabilizer codes known as
CSS codes.  These quantum codes have desirable properties especially
in the context of fault tolerant quantum computation.  Even though
some better bounds are known for CSS codes, such as tighter linear
programming bounds, it remained unclear whether they obey the quantum
Hamming bound. 

In this section, we will additionally assume that the alphabet size
$q$ is power of a prime.  We show that all CSS codes obey the quantum
Hamming bound when the alphabet size $q\ge 5$. In particular, we can
partially complement the results of Theorem~\ref{th:qhbCondTight} by
including the range $k+d > {(1-\delta)n}$, where $\delta
=2eq^{-2}$.

For the background, we mention that the CSS construction used here can
be found in~\cite[Theorem~9]{calderbank98} and $q$-ary versions in
\cite{grassl04} or \cite{ketkar06}.  Our proof takes advantage of an idea
that has been introduced in~\cite[Theorem~8]{ashikhmin99}.
\begin{lemma}\label{lm:cssAuxCodes}
Let $Q$ be an $[[n,k,d]]_q$ CSS code derived from a pair of classical
codes $C_1\subset C_2\subset\F_q^n$, where $C_i$ is an $[n,k_1]_q$ code.
Then $Q$ implies the existence of  $[n-k_1,k,\geq d]_q$ and $[k+k_1,k,\geq d]_q$ codes.
\end{lemma}
\begin{proof}
Since $C_1\subset C_2$, the generator matrices of $C_1$ and $C_2$ can
be put in the form
\begin{eqnarray*}
G_{C_1}= \left[\begin{array}{cc} I_{k_1}& P \end{array}\right] \quad
G_{C_2}=\left[\begin{array}{cc} I_{k_1}& P \\0_{k\times k_1} &
A\end{array}\right].
\end{eqnarray*}
Since $C_2$ is an $[n,k_1+k]_q$ code we can further transform  $G_{C_2}$ to 
\begin{eqnarray*}
G_{C_2}&=& \left[\begin{array}{ccc} I_{k_1}& P'&P'' \\0_{k\times k_1}
& I_k&A' \end{array}\right] = \left[\begin{array}{ccc} I_{k_1}&
\multicolumn{2}{c}{P} \\0_{k\times k_1} & I_k&A' \end{array}\right].
\end{eqnarray*}
The code generated by $\left[\begin{array}{ccc} 0_{k\times (n-k)/2} &
I_k&A' \end{array}\right]$ is in $C_2\setminus C_1$ and has a distance
$d$. Because the first $k_1$ coordinates are zero we can also view it
as an $[n-k_1,k,d]_q$ code. 
The codes $C_2^\perp \subset C_1^\perp$ have the parameters
$[n,n-k_1-k]_q$ and $[n,n-k_1]_q$ respectively. Reasoning similarly
with $C_2^\perp$ and $C_1^\perp$ we can show that there exists a
$[k_1+k,k,d]_q$ code. 
\end{proof}

\begin{proposition}\label{th:cssLargeQ}
Let $\mc{Q}$ be an $[[n,k,d]]_q$ CSS code with $k+d >
{(1-\delta)n}$ such that $\delta = 2eq^{-2} $ and $q$ a prime
power $\geq 5$. Then $\mc{Q}$ obeys the quantum Hamming bound.
\end{proposition}
\begin{proof}
Suppose that $\mc{Q}$ is derived from a pair of nested codes
$C_1\subset C_2 \subset \F_q^n$ with the parameters $[n,k_1]_q$ and
$[n,k+k_1]_q$, respectively.  These codes must satisfy
$\min\{\wt(C_2\setminus C_1), \wt(C_1^\perp\setminus C_2^\perp)\}=d$.

If $k+d =n-d+2$, then $\mc{Q}$ is an MDS code. Rains has shown that
every quantum MDS code is nondegenerate, see
\cite[Theorem~2]{rains99}; hence, the Hamming bound holds. Thus, we can
assume that $k+d\leq n-d+1$. The integrality
of $k+d$ implies that $k+d\geq \floor{(1-\delta)n}+1$.
By assumption, we also have 
$\floor{(1-\delta)n}+1\leq
k+d\leq n-d+1$, which implies $d\leq n-\floor{(1-\delta)n} =
\ceil{\delta n}$ and
\begin{eqnarray}
t=\floor{(d-1)/2} \leq \delta n/2.\label{eq:jRange}
\end{eqnarray}
By Lemma~\ref{lm:cssAuxCodes}, there exist classical codes $D$ and $D'$
with the parameters $[n-k_1,k,d]_q$ and $[k+k_1,k,d]_q$ respectively.
Since $D$ obeys the classical
Singleton bound, cf.~\cite[pg.~71]{huffman03}, we have
\begin{eqnarray}
n-k_1 &\geq& k+d-1. \label{eq:cS1}
\end{eqnarray}
In particular, if $k_1> n-k-d+1$, then $\mc{Q}$ cannot have a
distance $d$ and no $[[n,k,d]]_q$ code can be derived from such a
$C_1$ and $C_2$.  Further, $D$ obeys the classical Hamming bound,
see~\cite[pg.~48]{huffman03}; hence,
\begin{eqnarray}
q^{k}&\leq &\frac{q^{n-k_1}}{\sum_{j=0}^{t}\binom{n-k_1}{j}(q-1)^j}.\label{eq:cHB1}
\end{eqnarray}

Similarly, applying the classical Singleton and Hamming
bounds to $D'$, we respectively obtain
\begin{eqnarray}
k_1+k&\geq &k+d-1, \label{eq:cS2}\\ 
q^{k}&\leq &\frac{q^{k_1+k}}{\sum_{j=0}^{t}\binom{k_1+k}{j}(q-1)^j}.\label{eq:cHB2}
\end{eqnarray}
In particular, if $k_1<d-1$, there cannot exist an $[[n,k,d]]_q$ code.  From
equations~(\ref{eq:cHB1})~and~(\ref{eq:cHB2}) we obtain
\begin{eqnarray}
q^{2k}&\leq
&\frac{q^{n-k_1+k_1+k}}{\sum_{j=0}^{t}\binom{n-k_1}{j}(q-1)^j
\sum_{j=0}^{t}\binom{k_1+k}{j}(q-1)^j}. \nonumber 
\end{eqnarray}
which yields
\begin{eqnarray}
q^{k}&\leq
&\frac{q^{n}}{\sum_{i,j=0}^{t}\binom{n-k_1}{i}\binom{k_1+k}{j}(q-1)^{i+j}}.
\label{eq:cHB}
\end{eqnarray}
To prove that $\mc{Q}$ obeys the Hamming bound, it suffices to show
that the right hand side of (\ref{eq:cHB}) is less than the right hand
side of~(\ref{eq:qhb}); put differently, it suffices to show that 
$$ 
\sum_{j=0}^{t}\binom{n}{j}(q^2-1)^j \leq 
\sum_{i,j=0}^{t}\binom{n-k_1}{i}\binom{k_1+k}{j}(q-1)^{i+j}.
$$ 
If $n\leq 4$ and $k>0$, the quantum Singleton bound implies that $d\leq 2$, i.e., $t=0$
and the inequality holds. For $n\geq 5$ we shall prove an even stronger inequality, namely that 
\begin{eqnarray}
\sum_{j=0}^{t}\binom{n}{j}(q^2-1)^j \leq \sum_{j=0}^{t}\binom{n-k_1}{j}\binom{k+k_1}{j} (q-1)^{2j} \label{eq:qhbT}
\end{eqnarray}
holds term by term, $\binom{n}{j}(q^2-1)^j\leq
\binom{n-k_1}{j}\binom{k+k_1}{j}(q-1)^{2j} $. It clearly
holds for $j=0$. 
For $j>0$ we use the fact that $(n/j)^j\leq
\binom{n}{j} \leq (ne/j)^j$;  hence, it suffices to show that
\begin{eqnarray*}
\left(\frac{ne}{j}\right)^j(q^2-1)^j& \leq& \left(\frac{n-k_1}{j}\frac{k+k_1}{j}\right)^{j}(q-1)^{2j}.
\end{eqnarray*}
This is equivalent to showing that 
\begin{eqnarray}
\frac{ne}{j}(q+1)\leq\frac{n-k_1}{j} \frac{k+k_1}{j} (q-1).
\label{eq:termByTerm}
\end{eqnarray}
Notice that equality cannot hold in both~(\ref{eq:cS1})
and~(\ref{eq:cS2}). Indeed, if we have $k_1=n-k-d+1$
in~(\ref{eq:cS1}), then it follows that $k_1+k = n-d+1 \geq k+d$ as
$\mc{Q}$ is not MDS, tightening the inequality~(\ref{eq:cS2}).  If
$k_1+k=k+d-1$ in~(\ref{eq:cS2}), then this implies $n-k_1=n-d+1 \geq
k+d$, tightening the inequality~(\ref{eq:cS1}). It follows that 
$(n-k_1)(k+k_1)\ge (k+d)(k+d-1)\ge (1-\delta)n\,((1-\delta)n-1)$. 
Hence, to prove that
(\ref{eq:termByTerm}) holds it is enough to show
$$ej(q+1)\leq n(1-\delta)(1-\delta-1/n)(q-1).$$
By assumption $\delta=2 eq^{-2}$. By equation~(\ref{eq:jRange}), we have 
$j\leq t\leq \delta n/ 2$; thus, it remains to show that 
\begin{eqnarray*}
e^2q^{-2}(q+1) \leq  (1-2eq^{-2})(1-2eq^{-2}-1/n) (q-1).
\end{eqnarray*}
This inequality holds for $q=5$ and $n=5$. The left side of this inequality is
monotonically decreasing in $q$ while the right hand side is
monotonically increasing in $q$ and $n$; hence, the inequality holds for all 
$q\geq 5$ and $n\geq 5$. Consequently, we have shown that
inequality~(\ref{eq:qhbT}) holds for all $n$, and it follows that $\mc{Q}$ obeys the
quantum Hamming bound.  
\end{proof}

\begin{theorem}\label{th:cssAllqg5}
For $q\geq 5$ all $[[n,k,d]]_q$ CSS codes  obey the quantum Hamming bound.
\end{theorem}
\begin{proof}
Set $\delta = 2eq^{-2}$.  A CSS code obeys the quantum Hamming
bound by Theorem~\ref{th:qhbCondTight} if $k+d\leq
{(1-\delta)n}$, and by Proposition~\ref{th:cssLargeQ} if $k+d >
{(1-\delta)n}$.
\end{proof}

Other interesting bounds can be derived as a consequence of Lemma~\ref{lm:cssAuxCodes}.
For instance, an analogue of the Griesmer bound  is possible.
\begin{theorem}[Quantum Griesmer Bound for CSS Codes]\label{th:qGrmBnd}
An $[[n,k,d]]_q$ CSS code satisfies the following  bound:
\begin{eqnarray}
\frac{n+k}{2} &\geq & \sum_{i=0}^{k-1} \ceil{\frac{d}{q^i}}.
\end{eqnarray}
\end{theorem}
\begin{proof}
By Lemma~\ref{lm:cssAuxCodes} there exist $[n-k_1,k,d]_q$ and
$[k+k_1,k,d]_q$ codes. These codes  obey the classical
Griesmer bound, see \cite[Theorem~2.7.4]{huffman03}, hence we obtain
\begin{eqnarray*}
n-k_1  \geq   \sum_{i=0}^{k-1} \ceil{\frac{d}{q^i}} \mbox{ and }
 k+k_1 \geq \sum_{i=0}^{k-1} \ceil{\frac{d}{q^i}}.
\end{eqnarray*}
Combining the two inequalities proves the statement of the theorem.
\end{proof}
We can also show that a similar bound (though not exactly the same)
is applicable for linear quantum codes. Since $\ceil{d/q^i} \geq 1$ for $i>0$, 
we have $(n+k)/2 \geq d+k-1$ and we recover the quantum Singleton bound
as $n-k \geq 2d-2$. 
A very natural question would be if there are quantum codes that meet 
the quantum Griesmer bound. If $k=1$ (and $n-k$ even), then this essentially reduces to the
quantum Singleton bound and all $[[n,1,(n+1)/2]]_q$ quantum MDS codes meet this
bound. The interesting case is when $k\geq 2$. 
The $[[4,2,2]]_2$ code for instance meets this bound, 
it also meets the quantum Singleton bound. At this time we are not aware of other 
codes that meet the quantum Griesmer bound.

\begin{corollary}\label{co:tightSingBnd}
An $[[n,k,d]]_q$ CSS code with $d\geq q$ satisfies 
\begin{eqnarray}
\frac{n-k}{2} &\geq & d(1+1/q) - 2.
\end{eqnarray}
\end{corollary}
\begin{proof}
This is an easy consequence of Theorem~\ref{th:qGrmBnd}. Since $d\geq q$ we 
have 
\begin{eqnarray*}
\frac{n+k}{2} &\geq& d+d/q+\sum_{i=2}^{k-1}\ceil{\frac{d}{q^i}} \geq  d+d/q+k-2.
\end{eqnarray*}
Simplifying the above inequality yields the claim.
\end{proof}
Note that Corollary~\ref{co:tightSingBnd} is tighter than the quantum
Singleton bound.  Rains had shown that the binary quantum codes cannot
correct more than $\floor{(n+1)/6}$ errors \cite{rains99b}. A slightly stronger result
can be easily derived for CSS codes.
\begin{corollary}
An $[[n,k,d]]_2$ CSS code cannot correct more than 
$\floor{(n-k+1)/6}$ errors.
\end{corollary}
\begin{proof}
By  Corollary~\ref{co:tightSingBnd}, 
we have $(n-k)/2 \geq 3d/2-2$, which implies the claim. 
\end{proof}

\section{Conclusions}
In this paper we have shown that the quantum Hamming bound holds for all
CSS codes with alphabet greater than 5. We also have shown a slightly
weaker result for general quantum codes. 
Our results give
ample evidence for the conjecture that the quantum Hamming bound holds
for all quantum codes. However, there still remain some gaps.  The
major remaining open question is the status of $((n,K,d))_q$ quantum
codes which do not satisfy the conditions in
Theorem~\ref{th:qhbCondTight}.  Some special cases of interest are 
linear stabilizer codes and CSS codes of small alphabet $q\leq 4$.

\section*{Acknowledgment}
We thank Alexei Ashikhmin for correcting an erroneous remark in a previous version 
of this manuscript and an anonymous reader for pointing out a subtle
issue with the original formulation of Theorem~\ref{th:qhbCondTight}
for arbitrary quantum codes.  This research is supported by NSF Career Award CCF 0347310 
and NSF grant CCF 0622201. P.S. is also supported by grants from NSERC, MITACS and CIFAR.

\end{document}